\documentclass[preprint2]{aastex}
\usepackage{graphicx}

\newcommand{\kms}{\ensuremath{{\rm km \ s^{-1}}}}                   

\newcommand{\EtaC}{Eta Car}
\newcommand{\EtaA}{\ensuremath{\eta_\mathrm{A}}}
\newcommand{\EtaB}{\ensuremath{\eta_\mathrm{B}}}

\shorttitle{time evolution of Eta Carinae's colliding winds}
\shortauthors{Gull et al.}
\slugcomment{Draft version \today}
\begin{document}

\title{Imaging with {\it HST} the time evolution of Eta Carinae's colliding winds \altaffilmark{1}}

\author{Theodore R. Gull}
\affil{Code 667, Astrophysics Science Division,
Goddard Space Flight Center,
Greenbelt, MD 20771, USA; Theodore.R.Gull@nasa.gov}
\author{Thomas I. Madura and Jose H. Groh\affil{Max-Planck-Institut fur Radioastronomie, Auf dem Hugel 69, D-53121 Bonn, Germany}}
\and
\author{Michael F. Corcoran\altaffilmark{2}}

\affil{CRESST and X-ray Astrophysics Laboratory, Goddard Space Flight Center, Greenbelt, MD 20771, USA}

\altaffiltext{1}{Support for program 12013 was provided by NASA through a grant from the Space Telescope Science Institute, which is operated by the Association of Universities for Research in Astronomy, Inc., under NASA contract NAS 5-26555.}
\altaffiltext{2} {Universities Space Research Association, 10211 Wincopin Circle, Ste 500, Columbia, MD 21044}
\begin{abstract}
We report new {\it HST}/STIS observations that map the high-ionization forbidden line emission in the inner arcsecond of Eta Car, the first that fully image the extended wind-wind interaction region of the massive colliding wind binary. These observations were obtained after the 2009.0 periastron at orbital phases 0.084, 0.163, and 0.323 of the 5.54-year spectroscopic cycle. We analyze the variations in brightness and morphology of the emission, and find that blue-shifted emission ($-$400 to $-$200 \kms) is symmetric and elongated along the northeast-southwest axis, while the red-shifted emission ($+$100 to $+$200 \kms) is asymmetric and extends to the north-northwest. Comparison to synthetic images generated from a 3-D dynamical model strengthens the 3-D orbital orientation found by Madura et al. (2011), with an inclination $i \approx\ $ 138\degr, argument of periapsis $\omega \approx\ $ 270\degr, and an orbital axis that is aligned at the same PA on the sky as the symmetry axis of the Homunculus, 312\degr. We discuss the potential that these and future mappings have for constraining the stellar parameters of the companion star and the long-term variability of the system.

\end{abstract}

\defcitealias{hillier01}{H01}
\defcitealias{hillier06}{H06}
\defcitealias{gull09}{G09}
\defcitealias{madura10}{M10}
\defcitealias{mehner10}{Me10}
\defcitealias{madura11}{M11}
\section{Introduction}
Eta Carinae, one of the most luminous, variable objects in our Milky Way, is sufficiently close \citep[$D = 2.3 \pm 0.1$ kpc,][]{smith06} that we can study many of its properties throughout the electromagnetic spectrum. As noticed by \cite{damineli96}, the object exhibits a 5.54-year orbital period characterized by a lengthy high ionization\footnote{Low and high ionization are used here to describe atomic species with ionization potentials (IPs) below and above 13.6 eV, the IP of hydrogen.} state with  multiple high ionization forbidden lines  that disappear during months-long low ionization state  \citep{damineli08_period}. \EtaC\ is considered to be a massive, highly eccentric \citep[$e\sim0.9$,][]{corcoran05, nielsen05} binary consisting of \EtaA, a luminous blue variable (LBV), and \EtaB, a hot, less massive companion not directly seen, but whose properties have been inferred from its effects on the wind of \EtaA\ and the photoionization of nearby ejecta (\citealt{verner05,teodoro08}; \citealt[][hereafter Me10]{mehner10}; \citealt{gmo10, gnd10})

The total luminosity, dominated by \EtaA, is $\ge$\ 5$\times$10$^6$ L$_\odot$\ \citep{davidsonandhumph97}, with the total mass of the binary exceeding 120 M$_\odot$\ \citep[][hereafter H01]{hillier01}. Radiative transfer modeling of {\it HST}/STIS spatially-resolved spectroscopic observations suggests that \EtaA\ has a mass $\gtrsim 90 \ M_{\odot}$, and a stellar wind with a mass-loss rate of $\sim 10^{-3} M_{\odot} \ \mathrm{yr}^{-1}$ and terminal speed of $\sim 500 - 600 \ \mathrm{km \ s}^{-1}$ 
(\citealt{hillier01}; \citealt[][hereafter H06]{ hillier06}). 
Models of the observed X-ray spectrum require the wind terminal velocity of \EtaB\ to be $\sim 3000$ \kms\ with a mass-loss rate of $\sim$ 10$^{-5} M_\odot$ yr$^{-1}$ \citep{pc02}. The spectral type of \EtaB\ has been loosely constrained via modeling of the inner ejecta to be a mid-O supergiant 
(\citealt{verner05, teodoro08}; \citetalias{mehner10}).

3-D numerical simulations suggest that the wind of \EtaB\ strongly influences the very dense wind of \EtaA, creating a low-density cavity and {\it inner} wind-wind collision zone (WWCZ) \citep{pc02,okazaki08,parkin09}. The geometry and physical conditions of this inner region have been constrained from spatially unresolved X-ray \citep{henley08}, optical \citep{nielsen07,damineli08_multi}, and near-infrared \citep{gmo10,gnd10} observations.

In addition to the interaction between the two winds in the inner region (at spatial scales comparable to the semi-major axis length, $a\approx15.4$ AU = 0$\farcs$0067 at 2.3 kpc), the 3-D hydrodynamical simulations predict an {\it outer, extended}, ballistic WWCZ that stretches to distances several orders of magnitude larger than the size of the orbit 
(\citealt{okazaki08}; \citealt[][hereafter M10]{madura10}).  
Observational evidence for an extended WWCZ comes from the analysis of previous {\it HST}/STIS longslit observations 
(\citetalias{gull09}; \citetalias{madura10}; \citealt[][hereafter M11]{madura11})
which revealed spatially-extended forbidden line emission from low- and high-ionization species at $\sim$ 0\farcs1 to 0\farcs7 (230 to 1600 AU) from the central core. During the high state, [\ion{Fe}{2}] line emission  extends up to $\pm$500 \kms\ along the STIS slit, while [\ion{Fe}{3}] line emission extends to $-$400 \kms\ for STIS slit position angles close to 45\degr. Radiative transfer modeling of the extended [\ion{Fe}{3}] emission \citepalias{madura10,madura11} tightly constrains the orbital inclination, $i\approx138\degr$, close to the axis of inclination of the Homunculus, and the argument of periapsis 240\degr\ $\lesssim\omega\lesssim270$\degr\ in agreement with most researchers 
(\citealt{damineli08_period, gmo10,parkin09} and references therein).
 This constraint invalidates the claim by several groups (\citealt{falceta09,kashi09} and references therein) that periastron occurs on the near side of \EtaA\ ($\omega=90\degr$).

Here we report new {\it HST}/STIS observations, the first that fully map the inner arcsecond high-ionization, forbidden line emission of Eta Car. Maps of [\ion {Fe}{3}] $\lambda\lambda$4659.35, 4702.85\footnote{All wavelengths  are measured in vacuum.} and [\ion{N}{2}] $\lambda$5756.19 recorded in early phases following the 2009.0 periastron event show changes in the wind structures excited by FUV radiation from \EtaB. These results demonstrate that structural changes can be followed using specific forbidden lines, leading to increased knowledge about interacting wind properties, the parameters of the binary orbit and, most importantly, the stellar properties of \EtaB.

\section{Observations }

The {\it HST}/STIS mapping observations were obtained after the successful repair of STIS  during Service Mission 4. The first visit occurred in June 2009 ($\phi = 12.084$\footnote{All observations are referenced by cycle number relative to cycle 1 beginning 1948 February, following the convention introduced by \cite{gd04}. The phase $\phi$ is zeroed to JD2482819.8 $\pm$ 0.5 with period $P=2022.7 \pm 1.3$ days \citep{damineli08_period}.}) as an early release observation demonstrating the repaired-STIS capabilities (Program 11506 PI=Noll). The second and third  visits were scheduled in December 2009 ($\phi = 12.163$) and October 2010 ($\phi = 12.323$) under a CHANDRA/{\it HST} grant (Program 12013, PI Corcoran).

All observations were performed with the $52\arcsec \times 0\farcs1$ longslit. The strongest, most isolated, high-ionization forbidden emission lines from the inner and outer WWCZs  are  [\ion{Fe}{3}]  $\lambda\lambda$ 4659, 4702 and  [\ion{N}{2}]  $\lambda$5756 \citepalias{gull09}. The STIS gratings, G430M, centered at $\lambda$4706, and G750M, centered at $\lambda$5734, provide a spectral resolving power of about 8000.

\begin{figure*}[!ht]
\centering
\resizebox{1.0\hsize}{!}{\includegraphics{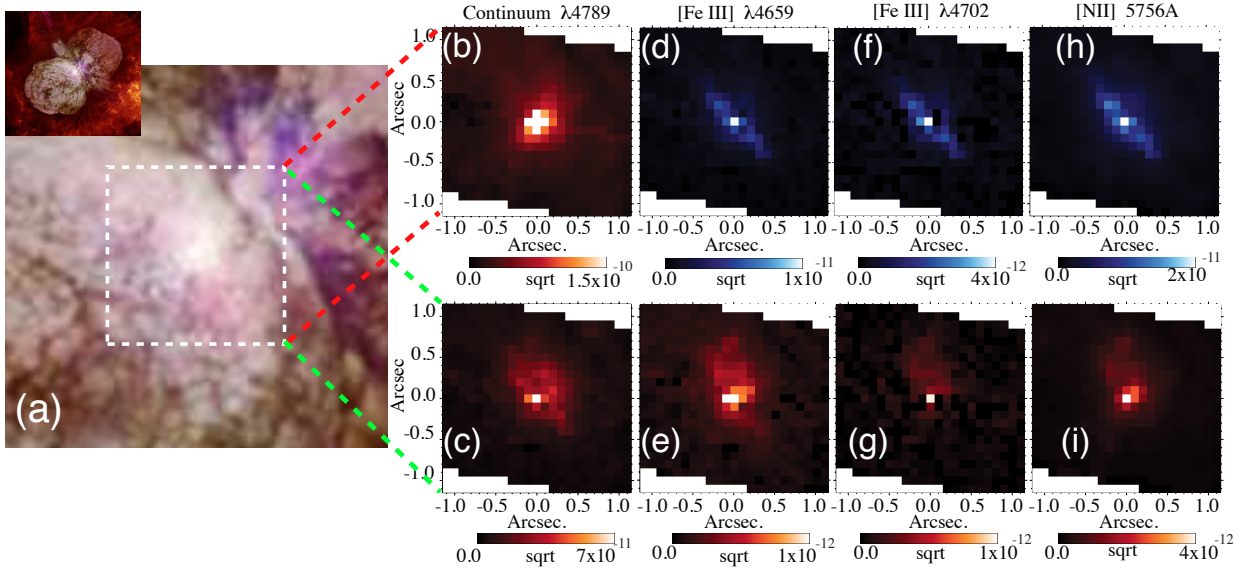}}
\caption{\label{mapping}{Comparison of red and blue images for isolated high-ionization forbidden lines from the $\phi=$12.084 observations (June 2009). (a)  {\it HST}/ACS image shows the 2\farcs2$\times$2\farcs2  box centered on Eta Carinae located within the 18\arcsec\ Homunculus as indicated in the small inset (HST archives). Strong continuum (b) has been subtracted from each forbidden emission map. [\ion{Fe}{3}]~$\lambda$4659 emission (c), integrated from $-$400 to $+$200 \kms, has a very different spatial distribution from the continuum. Blue images, extracted from $-$400 to $-$200 \kms\ for [\ion{Fe}{3}]~$\lambda$4659 (d), $\lambda$4702 (f) and [\ion{N}{2}]~$\lambda$5756 (h) are similar for each ion, as are red images extracted from $+$100 to $+$200 \kms\ for  [\ion{Fe}{3}]~$\lambda$4659 (e), $\lambda$4702 (g) and [\ion{N}{2}]~$\lambda$5756 (i). Images are displayed as sqrt(ergs cm$^{-2}\ $s$^{-1}$). North is up, and east is left.}}
\end{figure*}

Spatial mapping was accomplished with the standard STIS-PERP-TO-SLIT mosaic routine using the 52\arcsec$\times$0$\farcs$1 aperture with multiple 0$\farcs$1 offset position spacings centered on Eta Carinae. The size of the map, given limited foreknowledge of the extended forbidden emission structure, was adjusted with each visit based upon the anticipated {\it HST}/STIS longslit position angle (PA), pre-determined by the {\it HST} solar panel orientation. As buffer dumps impact the total integration time, only the central CCD rows, typically 64 (3\farcs2) or 128 (6\farcs4), were read out. The PAs for each visit were PA = 79\degr ($\phi=12.084$), -121\degr  ($\phi=12.163$), and -167\degr ($\phi=12.323$). Since a full spatial map was obtained during each visit, the PA has little effect on the results presented here (see Figures \ref{mapping} and \ref{feiii4659}).

The data were reduced with STIS GTO CALSTIS software. While data quality is similar to  previous {\it HST}/STIS observations of Eta Car obtained  from 1998 to 2004 
(\citealt{davidson05}; \citetalias{gull09}),
 the CCD detector has increased number of hot pixels, some bad columns, and increased charge transfer inefficiencies. Bright local continuum (Figure \ref{mapping}b) was subtracted from each pixel, isolating the faint forbidden line emission (Figures \ref{mapping}c-i, \ref{feiii4659}). Velocity channels were co-added to produce blue ($-400$ to $-200$ \kms), low-velocity ($-90$ to $+30$ \kms), and red ($+$100 to $+$200~\kms) images for each of the high-ionization forbidden lines (Figure~\ref{feiii4659}). Only the high velocity blue and red maps are sensitive to the wind wind interaction that we model in this present work.  The low velocity maps are dominated by slow-moving, extended ejecta produced in the 19th century eruptions, and so are not discussed in detail here. A refinement to the current model will include a screen of condensations to account for the low-velocity emission.

\section{Results} \label{results}

\subsection{Morphology and time evolution of the extended wind-wind collision \label{results1}}

For each phase, we compared velocity-separated images of [\ion{Fe}{3}]~$\lambda\lambda$4659, 4702 and [\ion{N}{2}]~$\lambda$5756, and found remarkable similarities in the blue and red images between the three emission lines (see Figure \ref{mapping} for June 2009, $\phi=$12.084). 
Hereafter we focus on the [\ion{Fe}{3}]  $\lambda$4659 emission, which cannot be formed by the primary star alone. Emission of [\ion{Fe}{3}] requires 16.2 eV photons from \EtaB,  plus thermal collisions at   electron densities approaching N$_e$~=~10$^7$~cm$^{-3}$ \citepalias{gull09,madura10,madura11}. By comparison, [\ion{N}{2}] emission is produced by 14.6 eV photons at electron densities approaching  N$_e$~=~3$\times$10$^7$~cm$^{-3}$. As the primary star, \EtaA, produces significant numbers of 14.6~eV photons \citepalias{hillier01}, [\ion{N}{2}] emission does not fully disappear during  periastron 
(\citealt{damineli08_multi}; \citetalias{gull09}).
 However, the red emission from [\ion{Fe}{3}] $\lambda$4659.35 can be contaminated by blue emission from [\ion{Fe}{2}] $\lambda$4665.75. Likewise, the red emission image of [\ion{Fe}{3}] $\lambda$4702.85 may be depressed by \ion{He}{1} $\lambda$4714.47 absorption. Hence, we examined the [\ion{N}{2}] maps to ensure little or no red high-ionization emission is present.

\begin{figure*}
\centering
\resizebox{.85\hsize}{!}{\includegraphics{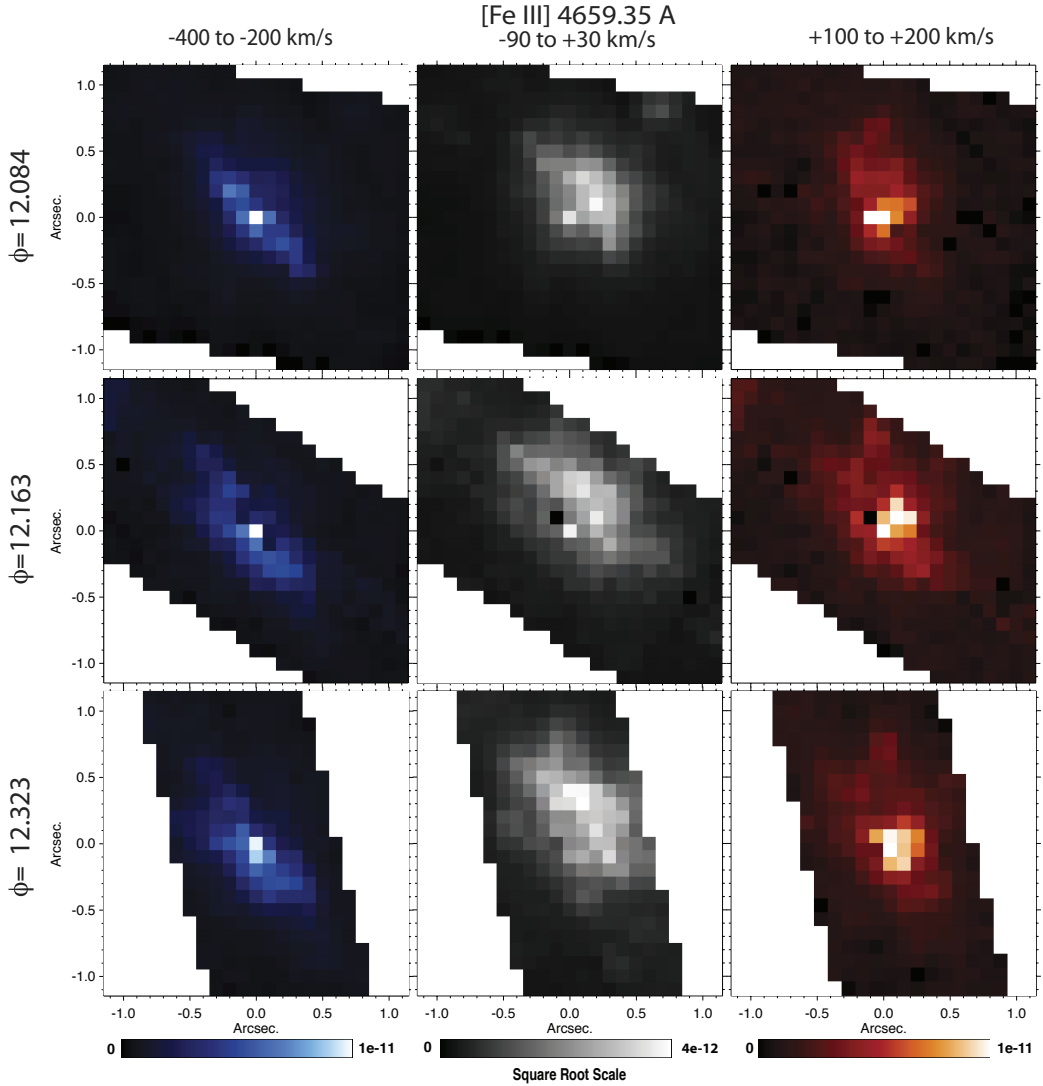}}
\caption{\label{feiii4659}{The changing shape of high-ionization [\ion{Fe}{3}]~$\lambda$4659 early in Eta Carinae's binary period. Top row: $\phi=$12.084. Middle row: $\phi=$12.163. Bottom row: $\phi=$12.323. Left column: blue emission ($-$400 to $-$200 \kms). Middle column: low-velocity emission ($-$90 to $+$30 \kms). Right column: red emission ($+$100 to $+$200 \kms). Gaps between the velocity intervals are purposefully excluded to show very separate velocity fields. The color bars show flux scaled by sqrt(ergs cm$^{-2}$s$^{-1}$.})}
\end{figure*}

Figure \ref{feiii4659} shows the time evolution of the blue, low-velocity, and red components of [\ion{Fe}{3}]~$\lambda$4659 at orbital phases $\phi = 12.084$, 12.163, and 12.323.
The morphology and geometry of the  extended [\ion{Fe}{3}]  $\lambda$4659 emission resolved by {\it HST}/STIS changes conspicuously as a function of velocity and time. The blue emission extends along the NE--SW direction, along $\mathrm{PA} \simeq 45\degr$, which is similar to what has been suggested from previous sparse HST/STIS long-slit observations obtained at different orbital phases across cycle 11 (G09, Me10, M10, M11). At $\phi=$12.084, the linear structure is nearly symmetrical about the central region, but at later phases becomes more diffuse, shifting to the S and SE. The red emission is fuzzier, asymmetric and extends primarily to the NNW at each phase. In contrast, the low-velocity structure is larger and extends diffusely northward. The low-velocity emission is heavily dominated by emission from the Weigelt blobs \citep{weigelt86} and a screen of fainter condensations \citepalias{mehner10}, located within the \EtaB\ wind-blown cavity and thusly obscuring the much fainter WWCZ contributions. While we describe the qualitative changes of the low-velocity component, we defer the detailed modeling of this equatorial emission to a future paper.

For discussion purposes, we now isolate the  central core (inner 0\farcs3$\times$0\farcs3) as representative of the inner WWCZ, and a time-variant extended ($>$0\farcs3$\times$0\farcs3) structure as representative of the outer WWCZ. These two regions have very different physical drivers. The central core exhibits X-ray \citep{pc02} and \ion{He}{1} emission, along with strong forbidden line emission. The outer WWCZ, expanding ballistically, is best traced by strong forbidden line emission. The spatially-extended blue and red emission components are thought to arise in the outer WWCZ of Eta Car \citepalias{gull09}, composed of material which was earlier part of the inner WWCZ, but over the past 5.5-year period streamed outward \citepalias{madura10,madura11}. While the primary wind is estimated to have a terminal velocity of $500-600$ \kms, the peak radial velocity component of the forbidden emission lines appears to be $\sim 400$ \kms. At terminal velocity, the outer WWCZ expands at 0\farcs25 per 5.5-year cycle, hence the current WWCZ, even at $\phi=$0.323, is within the 0\farcs3$\times$0\farcs3 core.

Both the central and extended structures brighten with phase, but they change differently. At $\phi=$12.084, the central core accounts for 1/3 of the flux, but brightens only thirty percent by $\phi=$12.323. The extended emission more than doubles in brightness by $\phi=$12.323. Brightening of the velocity components within the core and extended structures are likewise different. The brightness of the red component is nearly constant for both the core and the extended structure. The core blue component increases by seventy percent while the extended blue component doubles in brightness. The core low-velocity component increases only by fifty percent, but the low-velocity extended component triples in brightness and appears to shift further outward from the core.  We note that between $\phi =$ 12.163 and 12.323 the brightest low-velocity component shifts from the vicinity of Weigelt C, noted by \citetalias{mehner10}, to  Weigelt B and D.

These  brightness changes in the core and extended structures support a scenario in which the current WWCZ, namely the direct collision between the winds of \EtaA\ and \EtaB, is contained within the  0\farcs3 diameter core.  After each periastron passage, a new secondary-wind-blown cavity must form and expand outward. The cavity rapidly approaches a balance between the FUV flux of \EtaB\ and the cavity wall structure at critical density. However, the outer cavity wall is very thin, ionizes rapidly and drops in density allowing FUV radiation to pass outward into the much larger, ballistically expanding  outer cavity formed in the previous cycle. Within this cavity, the FUV photons encounter  dense walls of primary wind. The growth in brightness in the blue images, with little change in the red images, indicates expansion in the general direction of the observer. The larger increase in brightness of the low-velocity images shows where the FUV radiation escapes through the multiple cavities built up by the wind of \EtaB\ over many cycles.

\subsection{Comparison with a 3-D Dynamical Model}

Proper interpretation of the mapping observations requires a full 3-D dynamical model that accounts for the effects of orbital motion on the WWCZ. Here we use full 3-D Smoothed Particle
Hydrodynamics (SPH) simulations of Eta Car's colliding winds and
radiative transfer codes to compute the intensity in the [Fe III]
$\lambda$4659 line projected on the sky for a specified orbital
orientation \citepalias{madura10, madura11}. The numerical simulations were performed using the same 3-D
SPH code as that in \citet{okazaki08}  with identical parameters except for
 the mass loss rate of \EtaA, which we changed to 10$^{-3}$
M$_\odot$ yr$^{-1}$ \citepalias{hillier01, hillier06}. The two stellar winds in our
simulation are also taken to be adiabatic. In order to allow for a more
direct comparison to the {\it HST} observations, the computational domain is a
factor of ten larger than that of \citet{okazaki08} (i.e. $\pm$ 1600
$\mathrm{AU} \approx \pm$ 0\farcs7). Details on the radiative transfer calculations can
be found in \citetalias{madura10}, \citetalias{madura11}.

\begin{figure*}
\centering
\resizebox{0.9\hsize}{!}{\includegraphics{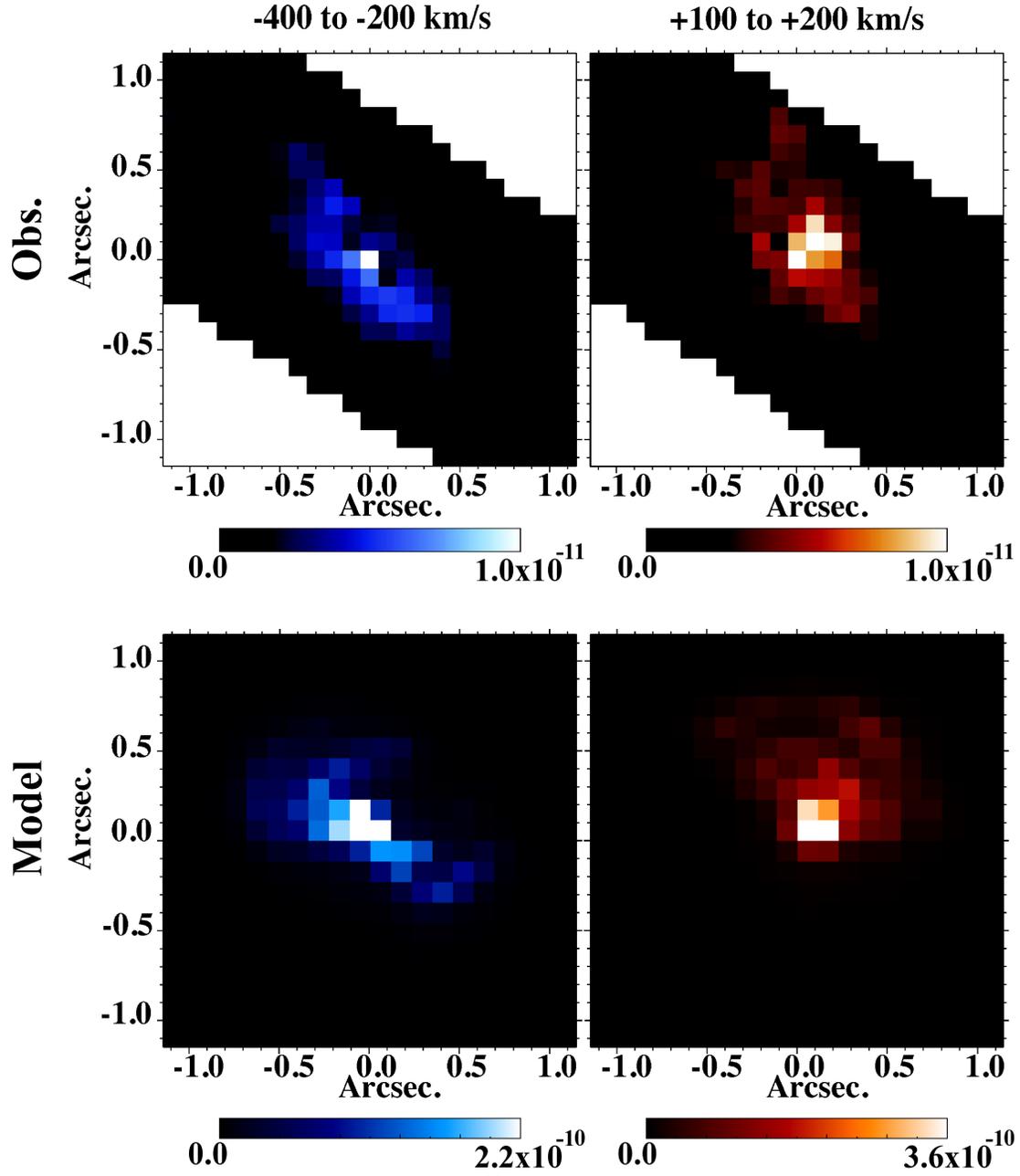}}
\caption{\label{model163} Comparison of $\phi=$12.163 blue and red components to 3-D dynamical model. Top row: Observed blue and red images. Bottom Row: 3-D SPH/radiative transfer images. Left column: $-$400 to $-$200 \kms. Right column: $+$100 to $+$200 \kms. Color display in all images is on a square root scale of ergs~cm$^{-2}$~s$^{-1}$. North is up.}
\end{figure*}

\begin{figure*}
\centering
\resizebox{0.9\hsize}{!}{\includegraphics{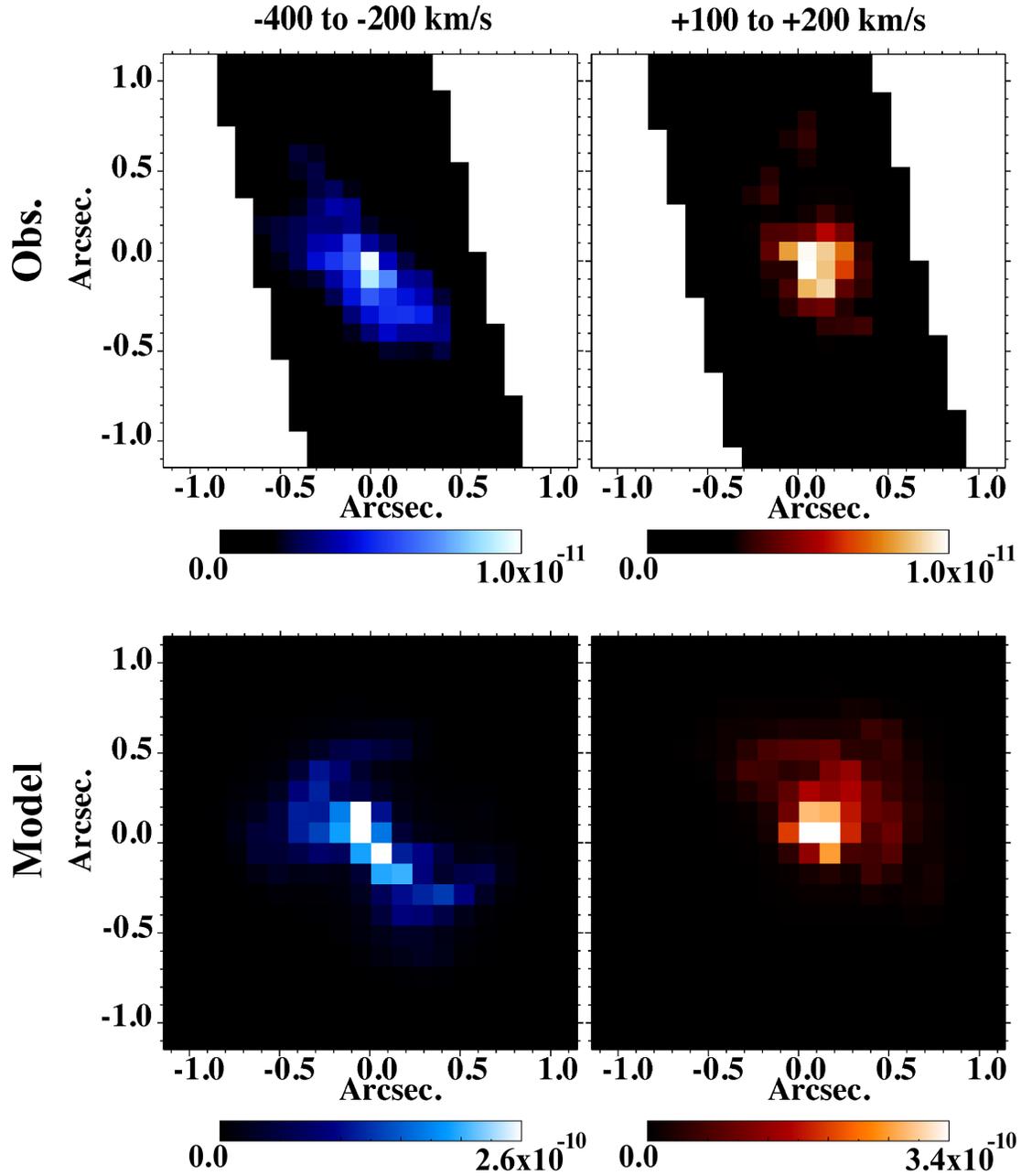}}
\caption{\label{model323} Comparison of $\phi=$ 12.323 blue and red components to 3-D dynamical model as in Figure \ref{model163}. Changes are subtle as \EtaB\ physically is close to the position of apastron; the ionization structure is primarily expanding.}
\end{figure*}

 Figures \ref{model163} and \ref{model323} compare the observed blue and red images at $\phi=$ 12.163 and 12.323 with those predicted by the model for the same velocity
intervals. For simplicity, the zero reference phase of the
spectroscopic cycle \citep{damineli08_multi}, is assumed to coincide with the zero reference phase of
the orbital cycle (i.e. periastron passage) in the 3-D SPH simulation. In a
highly-eccentric binary system like Eta Car, the two values should be within a few weeks, which will not affect the
overall conclusions \citep{gnd10}. The binary orbit is assumed to be
oriented with an inclination $i =$ 138\degr, argument of periapsis
$\omega =$ 270\degr, and an orbital axis that is aligned at the same
PA on the sky as the symmetry axis of the Homunculus,
312\degr\ \citep{davidson01}\footnote {\cite{davidson01} determined the Homunculus axis of symmetry to be tilted 42\degr\ into the sky plane. We refer the reader to \citetalias{madura11} for detailed discussion of the binary orbital inclination at 138\degr =180\degr-42\degr.}.

 The relatively compact central core produces little [\ion{Fe}{3}] emission  as densities in the WWCZ walls greatly exceed the critical density for efficient emission. The low-velocity maps, displayed on a flux scale similar to the scales for the blue and red images, would be blank while the observed low velocity emission, heavily dominated by flux from the Weigelt blobs and fainter slow-moving clumps, extends to the northwest. As mentioned in section \ref{results1}, we are refining the model to include such a screen, which will be a topic in a much more encompassing paper. Hence only the red and blue components, successfully replicated in this study, are presented in Figures \ref{model163} and \ref{model323}.

The spatial extent of the emission compares quite favorably between the
observations and the models  (Figures \ref{model163} and \ref{model323}), with the blue structures extending
projected distances of $\sim$
1\arcsec\ (2300 AU) along $\mathrm{PA} \sim$ 45\degr\ , and the red structures displaced to the NE of the core by $\sim$ 0\farcs1 to 0\farcs4 (230 to 1000 AU).   We display unreddened fluxes for the model structures due to known uncertainties of reddening. Model fluxes, reddened by $\approx$5--20 using typical *interstellar* reddening values for stars in the vicinity of Eta Car \citepalias{hillier01,mehner10} agree with the observations within a factor of a few. This discrepancy could arise due to uncertainties in the assumed stellar parameters of both stars, the reddening law  and atomic physics, or systematics in the radiative transfer and hydrodynamical modeling. However, reddening is highly variable across the Carinae complex. Moreover, reddening by dust in the Homunculus and within the extended core of \EtaC\ may change on very small scales. Hence we chose to display unreddened model fluxes in Figures \ref{model163} and \ref{model323}.

\section{Discussion} \label{disc}

This work represents the first time the extended WWCZ of a massive colliding wind binary system has been imaged using high-ionization forbidden emission lines. Spatial- and velocity-extended emission, recorded by individual {\it HST}/STIS longslit observations at various phases and PAs, provided impetus to expand 3-D models to simulate the wind dynamics leading to this emission.   Indeed, the initial 3-D dynamical model above produces red and blue images that are similar to those observed. From multiple longslit observations, 
\citetalias{gull09}, \citetalias{madura10} and \citetalias{madura11}
 demonstrated  that the binary orbit could be fully constrained in 3-D. The noticeable symmetry in velocity for observations taken at PA=38\degr\ \citepalias{gull09} is now reinforced by the spatial symmetry about the central core in the blue maps. Our modeling, of the observed maps suggests that the argument of periapsis must be closer to $\omega=$270\degr\ than 240\degr, thus further reinforcing the result that \EtaB\ is on the near side of \EtaA\ at apastron, with periastron passage on the far side 
 (\citealt{damineli97, pc02, okazaki08, parkin09}; \citetalias{gull09}, \citetalias{madura10}; \citetalias{madura11}). 

These and future spatial maps of \EtaC's high-ionization forbidden emission have the potential to determine the nature of the unseen companion star \EtaB. The mass-loss rate of \EtaA\ and ionizing flux of photons from \EtaB\ determine which regions of \EtaC's WWCZ are photoionized and capable of producing high-ionization forbidden line emission like the forbidden emission from Fe$^{++}$, due to 16.2 eV radiation.  Comparing this mass model loss rates and UV fluxes to those of stellar models for a range of O 
(\citealt{martins05}; \citetalias{mehner10}) 
and WR \citep{crowther07}
 stars would allow one to obtain a luminosity and temperature for \EtaB. Both the current model \citepalias{madura10, madura11} and previous individual {\it HST}/STIS longslit observations \citepalias{gull09} show major changes with orbital phase, especially near periastron. Mappings at multiple phases around periastron are therefore essential in order to determine when the FUV radiation from \EtaB\ becomes trapped in the dense wind of \EtaA\ and the extended high-ionization emission vanishes, and likewise when \EtaB\ emerges from \EtaA's wind and the extended emission returns.

This approach has a number of advantages over previous 1-D modeling efforts to constrain \EtaB's properties (\citealt{verner05}; \citetalias{ mehner10}), which probe the ionization structure of the Weigelt blobs. Such 1-D models make considerable assumptions about the physical conditions within the blobs and intervening material, leading to poor constraints on the luminosity of \EtaB. 

\EtaC\ is variable, not only on a 5.5-year period, but has a centuries-long history of variation, including two major eruptions \citep{davidsonandhumph97, humphreys08, smithfrew10}. These high-ionization forbidden emission lines are powerful tools for monitoring changes in the WWCZ, providing quantitative information on the properties of the individual binary components and changes thereof, including a historical record of the recent decade-long mass loss from the primary. Following this system will provide unique information on how a massive star, during the LBV stage, loses much of its mass on its way to becoming a supernova.

We sincerely thank G. Weigelt, S. Owocki, A. Damineli and A. Okazaki for many fruitful discussions and encouragements. TG acknowledges the hospitality of MPIR during his multiple visits. We thank the referee for insightful comments leading to an improved presentation.

\end{document}